\documentclass[useAMS,usenatbib,usegraphicx]{mn2e}
\title[The local star fraction-metallicity relation]
{
Understanding chemical evolution in resolved galaxies -- I\\
The local star fraction-metallicity relation
}
\author[Y.~Ascasibar et al.]
{
Y.~Ascasibar$^{\star 1}$, M.~Gavil\'{a}n$^{1}$, N.~Pinto$^{1}$, J.~Casado$^{1}$, F.~Rosales$^{2}$ and A.~I. D\'{i}az$^{1}$\\
$^{1}$ Departamento de F\'{i}sica Te\'{o}rica, Universidad Aut\'{o}noma de Madrid, Madrid 28049, Spain\\
$^{2}$ Instituto Nacional de Astrof\'{i}sica, \'{O}ptica y Electr\'{o}nica, Luis E. Erro 1, 72840 Tonantzintla, Puebla, Mexico
}

\date{\bf Draft version 3.0 (\today)}

\newcommand{\msun}{\ensuremath{\rm M_\odot}}

\newcommand{\mh}{\ensuremath{M_{\rm H}}}
\newcommand{\mgas}{\ensuremath{M_{\rm gas}}}
\newcommand{\mmetals}{\ensuremath{M_{\rm O}}}
\newcommand{\mstars}{\ensuremath{M_*}}

\newcommand{\zsn}{\ensuremath{\chi_{\rm SN}}}
\newcommand{\msn}{\ensuremath{m_{\rm SN}}}
\newcommand{\enriched}{\ensuremath{\varepsilon_{\rm w}}}

\newcommand{\avzsn}{\ensuremath{\bar\zsn}}
\newcommand{\avenriched}{\ensuremath{\bar\enriched}}

\newcommand{\avwind}{\ensuremath{\bar w_0}}
\newcommand{\avmetals}{\ensuremath{\bar\chi}}

\newcommand{\coeffgas}{\ensuremath{\Upsilon}}

\newcommand{\dd}{{\rm d}}
\newcommand{\deriv}[2]{\frac{\dd#1}{\dd#2}}

\newcommand{\eqref}[1]{(\ref{#1})}

\newcommand{\referee}[1]{{#1}}

\begin{document}

\maketitle

\begin{abstract}
This work studies the relation between gas-phase oxygen abundance and stellar-to-gas fraction in nearby galaxies.
We first derive the theoretical prediction, and argue that this relation is fundamental, in the sense that it must be verified regardless of the details of the gas accretion and star formation histories.
Moreover, it should hold on `local' scales, i.e. in regions of the order of 1~kpc.
These predictions are then compared with a set of spectroscopic observations, including both integrated and resolved data.
Although the results depend somewhat on the adopted metallicity calibration, observed galaxies are consistent with the predicted relation, imposing tight constraints on the mass-loading factor of (enriched) galactic winds.
The proposed parametrization of the star fraction-metallicity relation is able to describe the observed dependence of the oxygen abundance on gas mass at fixed stellar mass.
However, the `local' mass-metallicity relation also depends on the relation between stellar and gas surface densities.
\end{abstract}

\begin{keywords}
galaxies: fundamental parameters -- galaxies: abundances -- galaxies: evolution
\end{keywords}

\footnotetext[1]{E-mail: yago.ascasibar@uam.es}

%--------------------------------------------------------------------------
\section{Introduction}
%--------------------------------------------------------------------------

To a great extent, the evolution of any given galaxy through cosmic time is mainly defined by its gas accretion, star formation, and chemical enrichment histories.
All these aspects are deeply intertwined, and disentangling their complex interplay (and the physics behind it) is one of the long-lasting major goals of extragalactic astronomy.

For instance, it has long been known \citep{Lequeux+79} that the most luminous galaxies display higher metallicities than low-mass systems.
Although some of the earliest studies \citep[e.g.][]{Vigroux+87} found a weak correlation, subsequent works \citep[e.g.][among others]{Skillman+89, Vila-Costas&Edmunds92, Brodie&Huchra91, Zaritsky+94, Lee+03, Tremonti+04, Lamareille+04, Gallazzi+05, Perez-Montero+09_MZ} have convincingly shown that the gas-phase metallicity is a monotonically-increasing function of the stellar mass.
Albeit there is some scatter, the mass-metallicity (M-Z) relation has been established at all accessible redshifts \citep[e.g.][]{Savaglio+05, Erb+06, Maiolino+08}, and it appears to be mostly driven by internal processes rather than the environment \citep{Mouhcine+07, Hughes+13}.

The scatter of the M-Z relation has been found to correlate with the instantaneous star-formation rate (SFR), in the sense that galaxies of fixed stellar mass tend to show lower metallicities for higher SFR.
The existence of a `fundamental' M-Z-SFR relation, independent of redshift, that would explain the temporal evolution of the projection onto the M-Z plane in terms of the specific star formation rate at different cosmic epochs \referee{is still an open debate} \citep[cf.][]{Mannucci+10, Lara-Lopez+10, Yates+12, Perez-Montero+13, Andrews&Martini13, Lara-Lopez+13b, Lara-Lopez+13a, Zahid+13_FMR, Zahid+14_FMR, Salim+14}.

In addition, it has been shown \citep{Rosales+12, Sanchez+13} that there exists a `local' mass-metallicity relation between the spatially-resolved oxygen abundance and the stellar mass surface density, averaged over scales up to the order of one kpc.
However, no significant correlation with the star-formation surface density is observed \citep{Sanchez+13}.

Several theoretical interpretations have been proposed to account for these phenomenological relations, involving explanations based on a varying initial-mass function \citep[e.g][]{Koeppen+07}, inflow of pristine gas \citep[e.g.][]{Finlator&Dave08, Dave+10, Sanchez-Almeida+14}, selective outflows of metal-rich gas \citep[e.g][]{Garnett02, Tremonti+04, Dalcanton07, Kobayashi+07, Recchi+08, Spitoni+10, Peeples&Shankar11}, a systematic increase of the star-formation efficiency with stellar mass \citep[e.g][]{Brooks+07, Ellison+08, Calura+09, Vale-Asari+09, Gavilan+13, Hughes+13}, or a combination of several of these factors.

The gaseous component is of paramount importance in most of the proposed scenarios.
Recent observational studies, based on large samples of galaxies with 21-cm data and self-consistently derived metallicities, do indeed hint that the relation between metallicity and gas fraction, together with the dependence of gas fraction on stellar mass, are the main drivers of the observed M-Z and M-Z-SFR relations \citep{Hughes+13, Bothwell+13, Lara-Lopez+13a, Zahid+14}.
These studies show that the oxygen abundance is systematically higher for gas-poor galaxies, both for the galaxy population as well as at fixed stellar mass.

Since chemical elements are a by-product of star formation, and this is in turn a consequence of gas accretion, it seems quite natural that an underlying relation should exist between the amount of mass in stars, gas, and metals, either `locally' or integrated over the whole galactic extent.
Qualitatively, one may expect that the metallicity of gas-dominated systems (or regions) should be very low, because there have not been enough stars to enrich the interstellar medium, and that both stellar fraction and metallicity will grow in time as the galaxy evolves.

More quantitatively, exact analytical solutions have been derived under the instantaneous recycling approximation (IRA) for a closed-box model without any infall or outflow of gas \citep{Searle&Sargent72}, as well as for more general infall and outflow laws \citep{Edmunds90}.
In recent years, there is mounting evidence suggesting that galaxies probably grow near equilibrium, supporting a theoretical scenario often referred to as the `bathtub' or `gas-regulator' model  \citep{Finlator&Dave08, Recchi+08, Dekel+09, Bouche+10, Dave+12, Cacciato+12, Genel+12, Lilly+13, Feldmann13, Pipino+14}, where the accretion of gas is balanced by star formation and supernova-driven outflows.
It has also been shown that even sudden changes due to strong gas inflow episodes return to the analytical closed-box solution on relatively short timescales \citep{Dalcanton07}, and the role of such timescale on galaxy formation and evolution has been highlighted by \citet{Peng&Maiolino14}.

The present work investigates the physical origin of the observed correlation between the gas-phase oxygen abundance and the stellar-to-gas fraction, providing additional support to the idea that it is indeed more fundamental than the mass-metallicity relation on both local and global scales.
More precisely, we compare the theoretical prediction derived in Section~\ref{sec_model} \citep[cf.][]{Pipino+14} with a statistically-significant sample of publicly-available data, described in Section~\ref{sec_data}, that includes both integrated \citep[cf.][]{Zahid+14} and spatially-resolved spectroscopic observations.
The results of such comparison are presented in Section~\ref{sec_results}, where it is shown that the star fraction-metallicity relation also holds on kpc-scales, and it can account for the dependence of oxygen abundance on gas mass at fixed stellar mass \citep[cf.][]{Bothwell+13}.
In order to explain the observed mass-metallicity relation, a \emph{local} relation between the star fraction and the stellar surface density is also shown to exist, whose physical origin is still unclear.
Our main conclusions and prospects for future work in this direction are briefly summarized in Section~\ref{sec_conclusions}.

%--------------------------------------------------------------------------
\section{Theoretical prediction}
\label{sec_model}
%--------------------------------------------------------------------------

%---------------
\begin{figure}
\hfill\includegraphics[width=.45\textwidth]{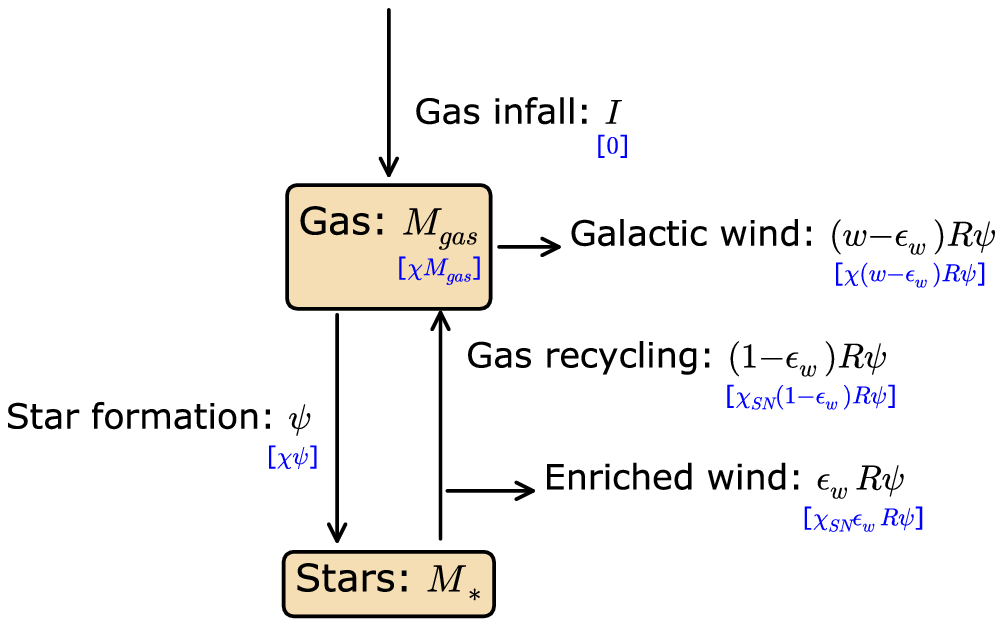}
\caption
{
Schematic representation of the main physical processes in our chemical evolution model.
Arrows indicate mass flow, and algebraic expressions correspond to the different terms in the system of differential equations discussed in Section~\ref{sec_model}.
Rates associated to the oxygen mass are shown in square brackets.
}
\label{fig_schema}
\end{figure}
%---------------

%--------------------------------------------------------------------------
\begin{table}
\begin{center}
\begin{tabular}{|l|l|}
\hline
Parameter    & Description \\
\hline
$\mgas(t)$    & Gas mass, in \msun.\\
$\mmetals(t)$ & Oxygen content of the gas, in \msun.\\
$\chi(t)$    & Gas-phase metallicity, $\mmetals(t)/\mgas(t)$.\\
$I(t)$       & Infall rate of external gas, in \msun~yr$^{-1}$.\\
$\psi(t)$    & Star formation rate, in \msun~yr$^{-1}$.\\
$R$          & Fraction of mass instantaneously returned\\
             & to the interstellar medium (ISM).\\
$\zsn(t)$    & Metallicity of the returned material (i.e. oxygen\\
             & abundance of the supernova ejecta).\\
$\enriched(t)$ & Fraction of the returned material that escapes\\
               & the galaxy before mixing with the ISM.\\
$w(t)R\psi(t)$ & Total galactic wind, in \msun~yr$^{-1}$.\\
$\mstars(t)$ & Stellar mass, in \msun.\\
\hline
$\avzsn(t)$      & Mass-weighted metallicity of the returned gas.\\
$\avenriched(t)$ & Oxygen mass-weighted fraction of the supernova\\
                 & ejecta that escapes the galaxy before mixing.\\
%                  & It sets the overall normalization of the\\
%                  & gas-fraction metallicity relation.\\
$\avwind(t)$     & Mass-weighted average intensity of the\\
                 & well-mixed component of the galactic wind.\\
%                  & relative to the returned gas mass, $w(t)-\enriched(t)$.\\
$\avmetals_w(t)$ & Mass-weighted metallicity of the well-mixed\\
                 & component of the galactic wind.\\
$\avmetals_*(t)$ & Mass-weighted stellar metallicity.\\
\hline
$\coeffgas(t)$   & Parameter that encapsulates the effect of all\\
                 & these processes on the maximum metallicity\\
                 & attainable when all the gas is turned into stars.\\
\hline
\end{tabular}
\end{center}
\caption
{
Physical interpretation of the different quantities appearing in our chemical evolution model.
}
\label{tab_parameters}
\end{table}
%--------------------------------------------------------------------------

In this section, we will derive a simple formula relating the gas-phase oxygen abundance to the ratio between stellar and gas mass.
The physical processes included in our chemical evolution model are illustrated in Figure~\ref{fig_schema}, and the meaning of all the different quantities defined below is succinctly summarized in Table~\ref{tab_parameters}.

Being a chemical element of primary origin, mainly produced in short-lived massive stars \citep[see e.g.][]{Wallerstein+97}, the evolution of the oxygen content can be followed under the instantaneous recycling approximation.
Under this assumption, the mass of diffuse gas in an isolated galaxy evolves as
\begin{equation}
\label{eq_gas}
\dot \mgas(t) = I(t) - \left[ 1-R + w(t)R\, \right]\, \psi(t)
\end{equation}
where some unspecified functions $I(t)$ and $\psi(t)$ represent the infall of external gas and the instantaneous star formation rate, respectively.
The constant $R$, denoting the fraction of gas returned to the interstellar medium, depends on the assumed IMF, and the presence of supernova-driven galactic winds is accounted for through a (time-dependent) mass-loading factor $w(t)$, defined in terms of the returned gas mass\footnote{Our prescription differs slightly from common practice, where a constant mass-loading factor $w'$ is expressed in units of the total amount of stars formed, i.e. the instantaneous star formation rate. In our notation, the usual definition corresponds to $w(t)=w'/R$.}.
At any given time, the mass in stars is given by
\begin{equation}
\label{eq_stars}
\dot \mstars(t) = (1-R)\, \psi(t)
\end{equation}
while the oxygen mass varies according to
\begin{eqnarray}
\label{eq_metals}
\nonumber \dot \mmetals(t)
\nonumber &=& \left[\, 1 - \enriched(t) \,\right]\, \zsn(t)\, R\, \psi(t) \\
\nonumber  && - \left[\, w(t)-\enriched(t) \,\right]\, \chi(t)\, R\, \psi(t) \\
 && -\ \chi(t)\, \psi(t).
\end{eqnarray}
The first term represents the injection of metals by the supernovae, where $\zsn(t)$ denotes the oxygen abundance of the ejecta as a function of time, and a fraction \mbox{$0 \le \enriched(t) \le 1$} of this enriched material may escape the galaxy before mixing with the surrounding gas (i.e. the enriched-wind scenario).
The second term corresponds to the rest of the galactic wind, assumed to have the same chemical composition as the interstellar medium (ISM); in particular, \mbox{$\chi(t) \equiv \mmetals(t)/\mgas(t)$}.
The mass-loading factor $w(t)$ includes both components of the galactic wind.
Finally, the third term accounts for the oxygen that is consumed by forming the next generation of stars.
This term usually appears in the literature multiplied by a factor $(1-R)$, to account (separately from the first term) for the fact that stars also return the original metals that were already present in the gas they were made from.
Although the difference is small in practice, we would like to advocate for a roughly constant oxygen \emph{abundance} of the ejecta, $\zsn$, rather than a constant \emph{yield} of freshly-produced oxygen, since the latter becomes unphysical as $\chi\to 1$.

Defining the quantities
\begin{eqnarray}
\avzsn(t) &\equiv& \frac{ \int_0^t \zsn(t')\, \psi(t')\, \dd t' }{ \int_0^t \psi(t')\, \dd t' }\\
\avenriched(t) &\equiv& \frac{ \int_0^t \enriched(t')\, \zsn(t')\, \psi(t')\, \dd t' }{ \int_0^t \zsn(t')\,\psi(t')\, \dd t' }\\
\avmetals_*(t) &\equiv& \frac{ \int_0^t \chi(t')\, \psi(t')\, \dd t' }{ \int_0^t \psi(t')\, \dd t' }\\
\avwind(t) &\equiv& \frac{ \int_0^t \left[\, w(t') - \enriched(t') \,\right]\, \psi(t')\, \dd t' }{ \int_0^t \psi(t')\, \dd t' }\\
\avmetals_w(t) &\equiv& \frac{ \int_0^t \chi(t')\, \left[\, w(t') - \enriched(t') \,\right]\, \psi(t')\, \dd t' }{ \int_0^t \left[\, w(t') - \enriched(t') \,\right]\, \psi(t')\, \dd t' }
\end{eqnarray}
equation~\eqref{eq_metals} can be explicitly integrated in time
\begin{eqnarray}
\nonumber \frac{ (1-R)\, \mmetals(t) }{ \mstars(t) }
\nonumber &=& \left[\, 1 - \avenriched(t) \,\right]\, \avzsn(t)\, R \\
\nonumber & & -\ \avmetals_w(t)\,\avwind(t)\, R \\
 & & -\ \avmetals_*(t)
\end{eqnarray}
and rewritten as
\begin{equation}
\label{eq_Zf}
\chi(t) = \frac
{ \left[\, 1 - \avenriched(t) \,\right] \avzsn(t)\, \frac{R}{1-R} \frac{\mstars(t)}{\mgas(t)} }
{ 1 + \coeffgas(t) \frac{R}{1-R} \frac{\mstars(t)}{\mgas(t)} }
\end{equation}
with
\begin{equation}
\coeffgas(t) \equiv \frac{ \avmetals_*(t) + \avmetals_w(t)\avwind(t)R }{ \chi(t)R }.
\label{eq_coeffgas}
\end{equation}
This expression has the same functional form as the ideal-regulator solution of \citet{Lilly+13}, and it is almost mathematically equivalent to that derived in \citet{Pipino+14} for the general case from a nearly identical set of equations.
As pointed out by these authors in connection with the specific star formation rate, we would like to stress that \emph{the relation between oxygen abundance and stellar-to-gas fraction described by equation~\eqref{eq_Zf} is largely insensitive to the details of the gas infall/outflow and star formation histories}.

In fact, all the dependence is encapsulated in the quantity $\coeffgas$.
Although this factor may in principle be any arbitrary function of time, completely different from one galaxy to another, its physical interpretation is fairly straightforward, and one may attempt to make an educated guess about its plausible range of reasonable values.

On the one hand, if the (enriched) wind parameters do not vary rapidly with time, $\avwind \approx w-\enriched$ will be approximately constant, and $\avmetals_w(t) \approx \avmetals_*(t)$ at all times.
On the other hand, a monotonically-increasing $\chi(t)$, i.e. a normal age-metallicity relation, implies
% \begin{equation}
$ 0 \le \frac{ \avmetals_*(t) }{ \chi(t) } \le 1 $,
% \end{equation}
with most relatively smooth functions $I(t)$ and $\psi(t)$ yielding values of order unity.
Thus, one expects that
\begin{equation}
\coeffgas \la \frac{ 1 }{ R } + \avwind
\end{equation}
is mostly tracing the fraction of stars exploding as supernovae (i.e. the IMF) and the presence of non-enriched galactic outflows.
Quantitatively, momentum conservation implies $w(t)\le \frac{ v_{\rm wind} }{ v_{\rm escape} }$; assuming $v_{\rm wind} \sim 2000$~km~s$^{-1}$, a escape velocity $ v_{\rm escape} > 20$~km~s$^{-1}$ implies that the average mass-loading factor should be in the range $ 0 \le \avwind \le 100 $.

Finally, the overall normalization of the relation is given by the enriched-wind parameter \avenriched\ and the adopted IMF, which sets the value of the recycled gas fraction
\begin{equation}
R = \frac
{ \int_{\msn}^\infty \left[\, m-m_r(m) \,\right]\, \phi(m)\, \dd m }
{ \int_{0}^\infty m\, \phi(m)\, \dd m }
\end{equation}
and influences the supernova yield
\begin{equation}
\zsn = \frac
{ \int_{\msn}^\infty \left[\, m_{\rm O}(m)-m_r(m) \,\right]\, \phi(m)\, \dd m }
{ \int_{\msn}^\infty \left[\, m-m_r(m) \,\right]\, \phi(m)\, \dd m }
\end{equation}
where $\msn\sim 8$~\msun\ is the mass above which all stars are assumed to instantaneously explode as type-II supernovae.

For a \citet{Kroupa} IMF, combined with the remnant mass and chemical composition of the ejecta computed by \citet{WoosleyWeaver}, one readily obtains $R \simeq 0.18$ and $\zsn \simeq 0.09$, almost independent of stellar metallicity.
Although many different IMFs have been proposed in the literature, their shapes above \msn\ are relatively similar.
Since the oxygen yield of massive stars is relatively well constrained, we do not expect that the actual value of \zsn\ is very different from our adopted value.
The value of $R$ is much more uncertain, as it depends critically on the shape of the IMF at low stellar masses (including the lower limit).
However, much of this dependence can be cancelled out by considering the quantity $\frac{R}{1-R}\mstars$ rather than the bare stellar mass\footnote{Note that the amount of massive stars, i.e. $\sim R\mstars$, is roughly what one measures when applying a given mass-to-light ratio, based on a certain IMF, to the observed luminosity.}.

To sum up, we predict that the metallicity of a galaxy must be approximately given by
\begin{eqnarray}
\nonumber
12 + \log({\rm O/H})
&\!\!=\!\!& 12 + \log(\, \frac{\mgas}{\mh} \frac{\chi}{16} \,) \\
% &\!\!\sim\!\!& (9.88-2.3\,\avenriched) + \log(\, \frac{ s }{ 1 + 1.35\, \coeffgas\, s } \,)
&\!\!\sim\!\!& 9.75+\log(1-\avenriched) + \log(\, \frac{ s }{ 1 + \frac{\coeffgas}{1.35} s } \,)
\label{eq_Z_s}
\end{eqnarray}
with $ s \equiv \frac{R}{1-R} \frac{\mstars}{\mh} $ and $ \mgas \sim 1.35\, \mh $ taking into account the contribution of helium.
In a weak-wind scenario ($\avenriched\ll 1$ and $\avwind\ll\coeffgas$), one expects $\coeffgas \la 1/R \sim 5.5$.
At low metallicities (high gas fractions), $12+\log({\rm O/H})$ increases linearly with $\log(s)$.
When most of the gas is converted into stars, the metallicity reaches an asymptotic value $12+\log({\rm O/H}) \to Z_0 \sim 9.75+\log(\frac{1.35}{\coeffgas})\sim 9.14$.
The transition between both regimes is expected to occur around $s\sim\frac{1.35}{\coeffgas}\sim 0.25$ or, equivalently, $\frac{\mstars}{\mgas}\sim 1.35(1-R)\sim 1.1$.

%--------------------------------------------------------------------------
\section{Observations}
\label{sec_data}
%--------------------------------------------------------------------------

\subsection{Integrated data}
%--------------------------------------------------------------------------

We have selected a large sample of star-forming galaxies with self-consistently measured values of the total atomic hydrogen mass, total stellar mass, and gas-phase metallicity.

We derive atomic hydrogen masses from radio observations of the 21-cm transition according to \citep[see e.g.][]{Wild52}
% A_21 = 2.85e-15 s^-1
% n1/n0 = 3*exp(-h\nu/kT_S) ~ 3*exp(-0.07 K/T_S) ~ 3
% photon rate = (3/4)*N_H*A_21
% S_\nu d\nu ~ S_\nu dv * \nu_0/c
% units '(2.36e5 solarmass/protonmass) * (.75*2.85e-15 s^-1) * h c/ 4 pi (1 Mpc)^2' 'Jy km/s'
\begin{equation}
 \frac{ M_{\rm HI} }{\msun}
 \simeq 2.36 \times 10^5\, \left(\frac{D}{\rm Mpc}\right)^2\, \frac{ S_{21} }{ \rm Jy~km~s^{-1} }
\end{equation}
where $D$ and $S_{21}$ denote the distance to the galaxy and the integrated flux density of the 21-cm line, respectively.
We assume that the atomic phase dominates the total hydrogen mass budget, i.e. $\mh \sim M_{\rm HI}$, neglecting the contribution of the molecular and ionized components.
Most of our objects (6984) have been detected by the Arecibo Legacy Fast ALFA (ALFALFA) survey \citep{Haynes+11}, and we consider 26 additional galaxies from the sample of extremely metal-poor (XMP) objects studied by \citet{Filho+13} in order to cover a broader metallicity range.

All our galaxies have optical spectroscopic data from the SDSS DR9 \citep{Ahn+12}, and they are classified as star-forming according to the \citet{Kewley+06} criterion.
Their stellar masses have been retrieved from the results of the MPA-JHU pipeline\footnote{\tt https://www.sdss3.org/dr9/algorithms/galaxy\_mpa\_jhu.php} available through the SDSS database.
They have been calculated for a Kroupa IMF, applying the Bayesian methodology and model grids described in \cite{Kauffmann+03} to the observed $ugriz$ fotometry of each object.

To estimate the nebular oxygen abundance, we resort to three different calibrators based on strong emission lines.
First, we consider the value reported by the Bayesian method implemented in the MPA-JHU pipeline \citep{Tremonti+04,Brinchmann+04}.
Then, we also consider the estimates based on the ratios $N2 \equiv \log\frac{\rm [NII]}{\rm H\alpha}$ and $O3N2 \equiv \log\frac{\rm [OIII]}{\rm H\beta}-N2$, according to the calibrations
\begin{equation}
12 + \log({\rm O/H}) = 9.07 + 0.79 \times N2
\label{eq_N2}
\end{equation}
and
\begin{equation}
12 + \log({\rm O/H}) = 8.74 - 0.31 \times O3N2
\label{eq_O3N2}
\end{equation}
proposed by \citet{Perez-Montero&Contini09}, using the line measurements from the MPA-JHU group %\footnote{Table {\tt galSpecLine}}
corrected for dust extinction by assuming an intrinsic H$\alpha$ to H$\beta$ ratio of~$2.8$ and a \citet{Miller&Mathews72} extinction curve with $R_V=3.2$ \citep[see e.g.][]{Haegele+08}.

\subsection{Resolved data}
%--------------------------------------------------------------------------

In principle, all the masses referred to in Section~\ref{sec_model} may correspond to any volume element.
Therefore, a `local' relation between the spatially-resolved oxygen abundance and the surface-density ratio $\Sigma_*/\Sigma_{\rm H}$ is also predicted to underlie the resolved $\Sigma_*$-Z relation discovered by \citet{Rosales+12} on scales of the order of one~kpc.

In order to test this prediction, we have compiled resolved HI, H$_2$, stellar mass, and metallicity observations for the galaxies NGC\,628 and NGC\,3184.
We have estimated stellar surface densities from the surface brightness of the 3.6-$\mu$m emission detected by the Spitzer Infrared Nearby Galaxies Survey \citep[SINGS\footnote{\tt http://irsa.ipac.caltech.edu/data/SPITZER/SINGS},][]{Kennicutt+03}, using a conversion factor \citep{Leroy+08}
\begin{equation}
 \frac{ \Sigma_* }{\rm \msun~pc^{-2} } = 280 \cos i\,\frac{ I_{3.6} }{\rm MJy~sr^{-1} }
 \label{eq_SigmaStars}
\end{equation}
where $i$ denotes the inclination angle of each galaxy (7 and 16 degrees for NGC\,628 and NGC\,3184, respectively).
For the neutral and molecular gas, we have used the moment-0 maps available from The HI Nearby Galaxy Survey \citep[THINGS\footnote{\tt http://www.mpia-hd.mpg.de/THINGS},][]{Walter+08} and the HERA CO-Line Extragalactic Survey \citep[HERACLES\footnote{\tt http://www.mpia-hd.mpg.de/HERACLES},][]{Leroy+09}, estimating the surface densities as
\begin{equation}
 \frac{ \Sigma_{\rm HI} }{\rm \msun~pc^{-2} } = 0.020 \cos i\, \frac{ I_{21} }{\rm K~km~s^{-1} }
 \label{eq_SigmaHI}
\end{equation}
for atomic hydrogen and
\begin{equation}
 \frac{ \Sigma_{\rm H_2} }{\rm \msun~pc^{-2} } = 5.5 \cos i\, \frac{ I_{\rm CO} }{\rm K~km~s^{-1} }
 \label{eq_SigmaH2}
\end{equation}
for the molecular component, where $I_{21}$ and $I_{\rm CO}$ correspond to the surface brightness of the 21-cm line and the $2\to1$ transition of carbon monoxide, respectively.
We have verified that the azimuthally-averaged profiles derived from our prescription are consistent with the results published in \citet{Leroy+08}.

Resolved observations of the gas-phase metallicity have been obtained from integral-field spectroscopic observations, carried out within the PPAK IFS Nearby Galaxy Survey \citep[PINGS\footnote{\tt http://xilonen.inaoep.mx/\textasciitilde frosales/pings/},][]{Rosales+10}.
Measurements of the oxygen and nitrogen forbidden lines, as well as hydrogen recombination lines entering the ratios $N2$ and $O3N2$ defined above have been compiled from the HII region catalog published by \citet{Sanchez+12}.
We have corrected the observed ratios for dust extinction by imposing H$\alpha$/H$\beta=2.8$ and estimated the current gas-phase oxygen abundance according to equations~\eqref{eq_N2} and~\eqref{eq_O3N2}.
The polynomial fit by \citet{Kewley&Ellison08} has then been used to convert the metallicity derived from the O3N2 indicator to the MPA-JHU calibration\footnote{In fact, the \citet{Kewley&Ellison08} conversion was derived for the \citet{Pettini&Pagel04} calibration, but the differences with \citet{Perez-Montero&Contini09} in the particular case of the O3N2 indicator are minor.}.

Since we are combining data from independent observational surveys, it is important to bear in mind their different spatial resolutions.
The HII regions defined in \citet{Sanchez+12} have a typical effective radius of the order of 3~arcsec for the two galaxies used in the present study ($\sim 150$~pc at a distance of $\sim 10$~Mpc).
This is comparable to the spatial resolution of the HI data \citep[robust-weighting synthesized beam major axis of $\sim 6$~arcsec according to][]{Walter+08}, but substantially worse than the 1.7~arcsec FWHM PSF of the IRAC camera at $3.6~\mu$m \citep{Fazio+04} and approximately one half of the final beam size of the CO observations \citep[13~arcsec;][]{Leroy+09}.
We have therefore downgraded the resolution of all our mass surface density maps to the latter value, plugging the average surface brightness within a radius $R=7$~arcsec ($\sim 700$~pc diameter) into equations~\eqref{eq_SigmaStars}, \eqref{eq_SigmaHI}, and \eqref{eq_SigmaH2}, and we made the assumption that the average metallicity does not vary appreciably on such small scales.
We have tested that, for galactocentric distances larger than $R$, our results are fairly robust with respect to the particular choice of this `resolution' parameter.

%---------------
\begin{figure*}
\includegraphics[width=\textwidth]{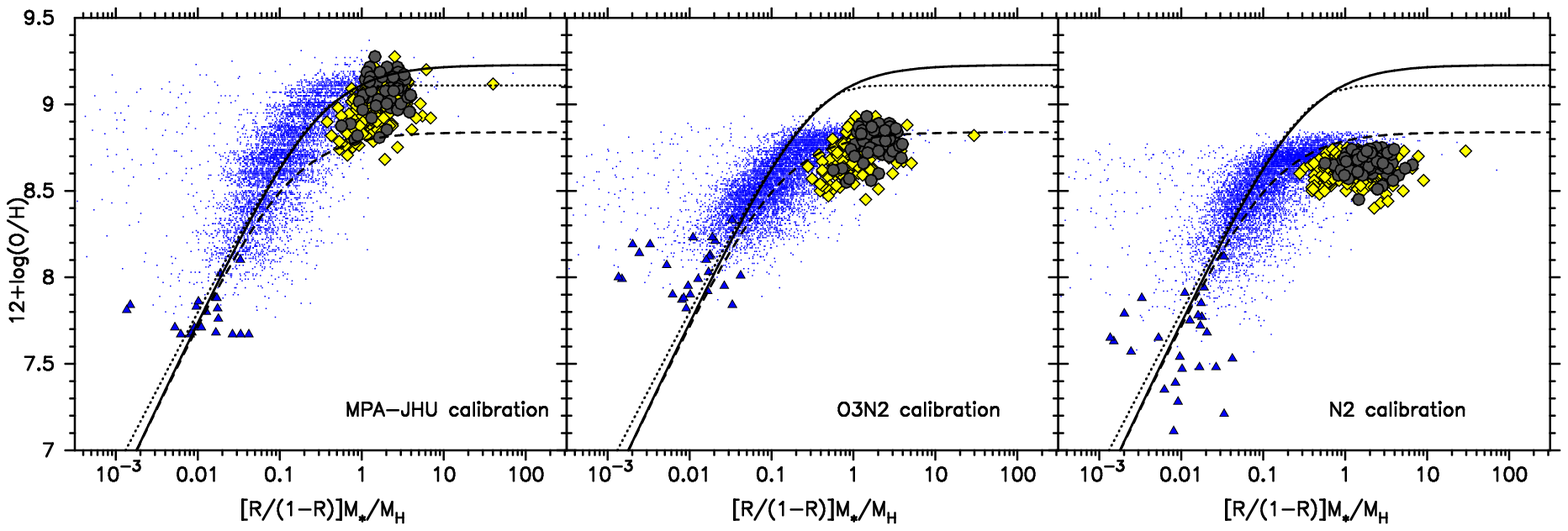}
\caption
{
Gas-phase metallicity as a function of the stellar-to-gas fraction.
Oxygen abundances derived from the MPA-JHU \citep{Tremonti+04}, O3N2 \citep{Perez-Montero&Contini09}, and N2 \citep{Perez-Montero&Contini09} calibrations are shown on the left, middle, and right panel, respectively.
Small dots represent SDSS data, and extremely-metal-poor galaxies from \citet{Filho+13} have been highlighted as triangles.
Resolved observations of NGC\,628 and NGC\,3184 are plotted as diamonds and circles, respectively.
Lines represent expression~\eqref{eq_Z_s} with $\avenriched=0$, $\coeffgas=4.5$ (solid) and $\coeffgas=11$ (dashed); dotted lines illustrate the results of \citet{Zahid+14}, i.e. equation~\eqref{eq_Zahid+14} with $Z_0=9.1$, based on the \citet{Kobulnicky&Kewley04} calibration.
}
\label{fig_Z_s}
\end{figure*}
%---------------

%---------------
\begin{figure*}
\includegraphics[width=\textwidth]{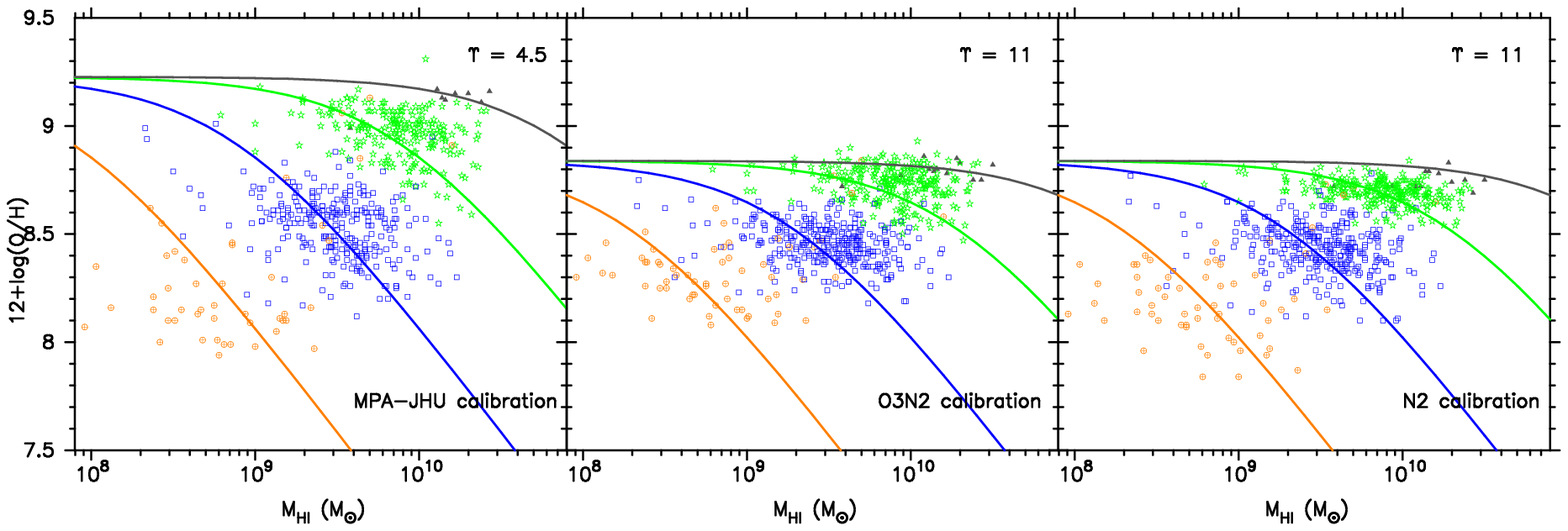}
\caption
{
Gas-phase metallicity as a function of gas mass.
Oxygen abundances derived from the MPA-JHU \citep{Tremonti+04}, O3N2 \citep{Perez-Montero&Contini09}, and N2 \citep{Perez-Montero&Contini09} calibrations are shown on the left, middle, and right panel, respectively.
Shapes denote stellar mass range: $0.9-1.1\times10^8~\msun$ (crosses), $0.9-1.1\times10^9~\msun$ (squares), $0.9-1.1\times10^{10}~\msun$ (stars), $0.9-1.1\times10^{11}~\msun$ (triangles).
Lines represent expression~\eqref{eq_Z_s} with $\avenriched=0$, $\coeffgas=4.5$ (for the MPA-JHU calibration) and $\coeffgas=11$ (for O3N2 and N2).
}
\label{fig_Z_g}
\end{figure*}
%---------------

%--------------------------------------------------------------------------
 \section{Results}
\label{sec_results}
%--------------------------------------------------------------------------

The relation between gas-phase oxygen abundance and stellar-to-gas fraction is depicted in Figure~\ref{fig_Z_s}.
Both integrated data (i.e. SDSS and XMP galaxies) as well as resolved observations of the stellar and total hydrogen column densities in NGC\,628 and NGC\,3184 describe a fairly tight relation, well described by equation~\eqref{eq_Z_s}.

In principle, the parameter $\coeffgas$ could vary from one galaxy to another, or even within different regions of the same galaxy, due to the local efficiency of galactic winds and/or the details of the infall and star formation histories, which set the ratio between the mass-weighted stellar metallicity $\avmetals_*(t)$ and the instantaneous gas-phase abundance $\chi(t)$.
Our results, however, show that such variations must be relatively mild, given that the observed relation can be described with a single value of \coeffgas\ over more than three orders of magnitude in stellar fraction.
The best-fitting value depends somewhat on the adopted metallicity calibration, ranging from $\coeffgas = 4.5$ for the MPA-JHU prescription \citep{Tremonti+04} to $\coeffgas \sim 11$ for the O3N2 and N2 calibrations of \citet{Perez-Montero&Contini09}.

A similar relation has recently been found by \citet{Zahid+14} for a sample of galaxies at $z<1.5$, including SDSS spectra,
\begin{equation}
 12 + \log({\rm O/H}) = Z_0 - \log( 1 - e^{-\frac{\mstars}{\mgas}} )
 \label{eq_Zahid+14}
\end{equation}
where $Z_0=9.1$, based on the \citet{Kobulnicky&Kewley04} calibration.
This relation, shown as a dotted line in Figure~\ref{fig_Z_s}, is perfectly consistent with expression~\eqref{eq_Z_s} with $\avenriched=0$ and $\coeffgas\approx 6$.
Although the proposed functional forms are quite different, the expressions $A(x)=\frac{x}{1+x}$ and $Z(x)=1-e^{-x}$ have a very similar qualitative behaviour.

The scatter around the best-fitting relation is relatively low, of the order of $0.21$, $0.13$, and $0.15$~dex for MPA-JHU, O3N2, and N2, respectively.
These numbers are slightly larger (about 20 per cent) than those obtained for a similar analysis of the mass-metallicity relation, and they are largely driven by gas-rich systems.
In the gas-poor regime (i.e. $\mstars>\mgas$; $s>\frac{R}{1-R}=0.22$), the scatter drops to $0.17$, $0.10$, and $0.11$~dex, respectively, becoming comparable to or lower than that of the mass-metallicity relation.
It would be interesting to understand whether the additional scatter at low metallicities is physical -- due to e.g. more stochastic (i.e. `bursty') accretion, star formation, and/or winds -- or it arises from a larger statistical uncertainty in the metallicity estimates.
In addition, it may also be worth mentioning that the gas distribution is often more extended than the star-forming region, and it is certainly larger than the diameter of the SDSS fibre within which the oxygen abundance is measured.
Although our model can be applied to any region where the instantaneous mixing approximation is valid, this spatial mismatch, combined with the measurement uncertainties in the gas masses, must contribute to the observed scatter.

For chemically young systems (i.e. low stellar fractions), the theoretical prediction of a linear relation is extremely robust: the oxygen mass is simply proportional to the stellar mass.
Including the factor $\frac{R}{1-R}$ in the definition of the stellar-to-gas fraction, the normalization depends only slightly on the assumed IMF, and it is almost entirely set by the amount of metals that mix with the interstellar medium, given by the product $(1-\avenriched)\zsn$.
Of course, the enriched-wind parameter $\avenriched$ and the supernova oxygen yield \zsn\ are absolutely degenerate.
For the adopted yield \citep{WoosleyWeaver}, our results strongly favour a scenario where supernova ejecta are quickly mixed with the surrounding gas, leaving little room for selective mass loss through metal-enriched winds.
In fact, the metallicities measured in these systems tend to be, on average, \emph{above} the predicted relation, perhaps hinting towards selection effects (e.g. minimum line intensities and/or equivalent widths) or a possible systematic bias of the calibrations that is more severe at low metallicities (e.g. a non-linear N2 or O3N2 calibration).

Concerning a well-mixed galactic wind, the best-fitting values of $\coeffgas$ reinforce the conclusion of \citet{Gavilan+13} that massive outflows are not required in order to reproduce the chemical properties of gas-rich galaxies.
Here we may even constrain the maximum value of the wind mass-loading factor in more evolved systems.
First, let us adopt $\avmetals_*(t) \approx 0.8 \chi(t)$ as a representative value \citep[see e.g.][for a detailed discussion of the stellar metallicity in NGC\,628]{Sanchez-Blazquez+14}, so that equation~\eqref{eq_coeffgas} yields $\coeffgas=4.5$ for $\avwind=0$ and $R=0.18$.
Assuming that the amount of gas expelled by the galactic wind is always proportional to the instantaneous star formation rate, a mass-loading factor $\avwind\ge17$ (or, in terms of the total stellar mass produced, $w'=\avwind R\ge3$) would result in $\coeffgas=4.5(1+w')\ge18$, and therefore an asymptotic value of $12+\log({\rm O/H})\le8.6$.
Such a low value severely underestimates the measurement inferred from the MPA-JHU calibration, and it lies somewhat below the observations based on the N2 and O3N2 indicators.
Note that, in such a scenario, supersolar (or even solar) metallicities should never be observed\footnote{It is also important to bear in mind at this point that neither the O3N2 nor (especially) the N2 calibrations are valid in the supersolar regime \citep[see e.g.][]{Perez-Montero&Contini09}.}.

From our point of view, other relations, such as the connection between oxygen abundance and HI mass observed by \citet{Bothwell+13}, are a consequence of the more fundamental star fraction-metallicity relation described by expression~\eqref{eq_Z_s}.
As can be readily seen in Figure~\ref{fig_Z_g}, the observed data may be quantitatively reproduced with $\avenriched=0$ and $\coeffgas=4.5$ for the MPA-JHU calibration and $\coeffgas \sim 11$ for the O3N2 and N2 prescriptions in \citet{Perez-Montero&Contini09}.

We must note, though, that the theoretical relation may be somewhat steeper than the observations, although the effect seems to be less severe than the discrepancy reported by \citet{Bothwell+13}.

On the one hand, the logarithmic slope derived from equation~\eqref{eq_Z_s} varies with stellar fraction as
\begin{equation}
 \deriv{ \log({\rm O/H}) }{ \log\mh } = -\frac{ 1 }{ 1+\frac{\coeffgas}{1.35} s }
\end{equation}
taking values between $-0.6$ and $-0.4$ for $4.5<\coeffgas<11$ and $s=\frac{R}{1-R}=0.22$ (i.e. $\mh=\mstars$), which roughly corresponds to the transition in the star fraction-metallicity relation.
A slope of $-1$, corresponding to a `pure dilution' scenario, is only attained asymptotically at very low stellar fractions.

On the other hand, the slope defined by the observational data is rather sensitive to the uncertainties in the determination of metallicities, stellar masses, and gas masses, as well as to the details of the respective mass functions.
Although a rigorous statistical study lies beyond the scope of the present work, Figure~\ref{fig_Z_g} suggests that the observed relation is compatible with equation~\eqref{eq_Z_s} over the mass range $10^8\le \mstars/\msun \le 10^{11}$, well covered by SDSS galaxies.
In particular, we do observe that the logarithmic slope of the gas mass-metallicity relation at fixed stellar mass seems to increase (i.e. become shallower) with stellar mass, in agreement with the theoretical expectation but at variance with the constant value of $-0.15$ observed by \citet{Bothwell+13}.

%---------------
\begin{figure}
\includegraphics[width=8cm]{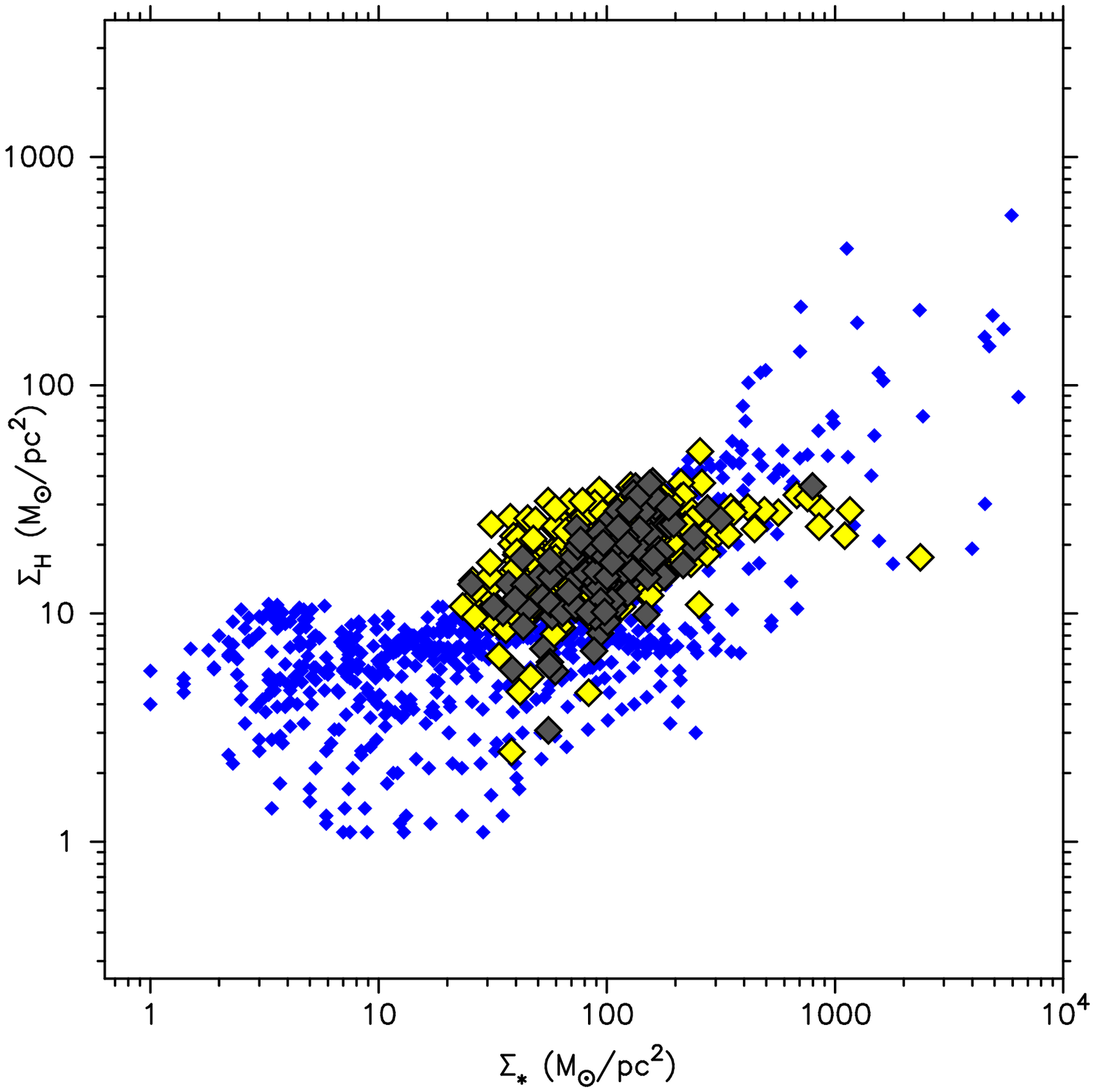}
\caption
{
Relation between stellar and gas surface density.
Resolved observations of NGC\,628 and NGC\,3184 are plotted as diamonds and circles, respectively.
Crosses show the values corresponding to the radial annuli defined by \citet{Leroy+08} for a sample of nearby galaxies.
}
\label{fig_Sigma_star}
\end{figure}
%---------------

Concerning the mass-metallicity relation, we support the interpretation \citep[see e.g.][and references therein]{Zahid+14} that it is a consequence of the universal metallicity-star fraction relation \emph{and} galactic `downsizing', understood as a systematic increase in stellar fraction with stellar mass.
Our results show that the star fraction-metallicity relation is verified on \emph{local} scales; therefore, the existence of a `local' mass-metallicity relation \citep{Rosales+12, Sanchez+13} hints that there should also exist a correlation between stellar fraction and surface density.

In order to verify that statement, we plot in Figure~\ref{fig_Sigma_star} the stellar and gas surface densities of our individual regions in NGC\,628 and NGC\,3184 together with the radial profiles, defined in elliptical annuli, obtained by \citet{Leroy+08} for a sample of nearby galaxies (including NGC\,628 and NGC\,3184).
The relation seems to be non-linear, but there is a clear trend towards more chemically-evolved systems at higher surface densities, with the transition between gas-rich and gas-poor regions occurring at a characteristic stellar surface density of the order of $\sim 20$~\msun~pc$^{-2}$.
The Spearman rank-order correlation coefficient is $0.73$, implying a negligible p-value (of the order of $10^{-175}$) for the data being uncorrelated.

In contrast to the star fraction-metallicity relation, the connection with the stellar surface density is entirely determined by the rate at which the gas is accreted into the galaxy (or region) and converted into stars, i.e. the functions $I(t)$ and $\psi(t)$.
In our opinion, this is the second fundamental relation that underlies chemical evolution, and therefore understanding its physical origin is of the utmost importance.

%--------------------------------------------------------------------------
 \section{Discussion and conclusions}
\label{sec_conclusions}
%--------------------------------------------------------------------------

There is mounting evidence that all the observable properties of normal galaxies, including their spectra, can be described in terms of one single free parameter, which may be interpreted as a quantitative indicator of the evolutionary state of the system \citep[see e.g.][and references therein]{Disney+08, Ascasibar&Sanchez-Almeida11}.
Physically, two such indicators are the stellar fraction and the gas metallicity: starting from zero, these two quantities increase as the gas is converted into stars, and they may only decrease as a consequence of pristine gas accretion.

In agreement with previous theoretical work \citep[e.g.][among many others]{Edmunds90, Pipino+14}, we find that the relation between star fraction and gas-phase oxygen abundance is rather insensitive to the details of the infall and star formation histories.
In gas-dominated systems, the amount of metals is approximately proportional to the stellar mass, and therefore the gas metallicity is a linear function of the star fraction.
In gas-poor (more evolved) objects, the gas-phase oxygen abundance reaches a plateau that represents an equilibrium between gas flows, star formation, and stellar yields.

Although the presence of strong galactic winds may have an impact on such equilibrium value, we find that the star fraction-metallicity relation can be described by expression~\eqref{eq_Z_s} for a broad class of scenarios.
All the particular details of the evolutionary history are encapsulated into a single coefficient, $\coeffgas$, whose definition~\eqref{eq_coeffgas} has a straightforward physical interpretation.

We thus conclude that the star fraction-metallicity relation is fundamental, in the sense that all galaxies (\emph{or individual regions within galaxies}) are expected to follow it, unless any of the basic assumptions of the model (e.g. perfect gas mixing) turned out to be a bad approximation.
In particular, equation~\eqref{eq_Z_s} is expected to hold at all times.
Observations of the mass-metallicity relation at different redshifts \citep[e.g.][]{Zahid+14} suggest that the relation is indeed approximately invariant over the range $0<z<1.6$.

In the present work, the comparison of the theoretical prediction with a a set of observational data including both integrated ($\sim7000$ objects with SDSS+ALFALFA data) and resolved spectra ($\sim 400$ H{\sc ii} regions from the galaxies NGC\,628 and NGC\,3184) suggests that:

\begin{enumerate}

\item Both integrated and resolved observations display a well defined relation between gas-phase metallicity and stellar-to-gas ratio.

\item Expression~\eqref{eq_Z_s} provides an accurate description of the observed trend.

\item For the adopted IMF \citep{Kroupa} and supernova yields \citep{WoosleyWeaver}, the fraction of metals directly ejected to the intergalactic medium without mixing with the galactic ISM (i.e. the mass-loading factor of \emph{enriched winds} \enriched) is consistent with zero.
Higher values are tightly constrained by the amount of metals present in gas-rich systems.

\item Overall, the star fraction-metallicity relation is consistent with a single value of $\coeffgas$ for all the observed galaxies and regions, covering more than three orders of magnitude in stellar-to-gas ratio.
However, the best-fitting value depends on the adopted metallicity calibration, ranging from $\coeffgas=4.5$ for MPA-JHU \citep{Tremonti+04} to $\coeffgas \sim 11$ for O3N2 or N2 \citep{Perez-Montero&Contini09}.

\item These values of \coeffgas\ imply an efficiency of the well-mixed galactic wind of the order of $\avwind R\sim 3$~times the instantaneous star formation rate for the O3N2 and N2 calibrations.
For the MPA-JHU pipeline, our results favour a scenario where massive outflows are not required in order to reproduce the chemical properties of nearby galaxies \citep[cf.][]{Gavilan+13}.
A detailed comparison of gas-phase and mass-weighted stellar metallicities would be extremely useful in order to provide more accurate information on the total amount of metals that escape from the galaxy.

\item We argue that the relation between star fraction and metallicity underlies other correlations, such as the dependence of oxygen abundance on gas mass at fixed stellar mass \citep{Bothwell+13} or the mass-metallicity relation.

\item However, the existence of a `local' mass-metallicity relation \citep{Rosales+12, Sanchez+13} \emph{also} requires a relation between star fraction and stellar surface density.
Such relation is indeed observed, although its physical origin is still unclear.

\end{enumerate}

In the proposed scenario, the star fraction-metallicity relation arises naturally, as both quantities behave as `clocks' that measure the evolutionary state of a given galaxy or region.
However, it is only one of the three fundamental ingredients that regulate galaxy formation and (chemical) evolution.
The other two are, of course, the accretion of external gas and its conversion rate into stars, whose interplay results in a (local) relation between star fraction and stellar mass (surface density), and therefore between the latter and the gas-phase metallicity.

Further work will be devoted to investigate the physical processes that underlie this relation by setting the functions $I(t)$ and $\psi(t)$.
If they depended on a single degree of freedom, there would be a one-to-one correspondence between all the observable properties of a galaxy; otherwise, one would expect an average trend, but also 'residual' correlations due to the existence of additional parameters (e.g. total mass, surface density, angular momentum, infall time...).
In this respect, establishing whether the specific star formation rate (density) is related to the offset with respect to the (local) mass-metallicity relation \citep{Salim+14} or not \citep{Sanchez+13} would provide a crucial constraint for chemical evolution models.

%--------------------------------------------------------------------------
 \section*{Acknowledgments}
%--------------------------------------------------------------------------

This research has made use of NASA's Astrophysics Data System, as well as public data from THINGS \citep[`The HI Nearby Galaxy Survey';][]{Walter+08}, HERACLES \citep[the `HERA CO-Line Extragalactic Survey';][]{Leroy+09}, PINGS \citep[the `PPAK IFS Nearby Galaxy Survey';][]{Rosales+10}, ALFALFA \citep[the `Arecibo Legacy Fast ALFA' survey;][]{Haynes+11}, and the Sloan Digital Sky survey.
Funding for SDSS-III has been provided by the Alfred P. Sloan Foundation, the Participating Institutions, the National Science Foundation, and the U.S. Department of Energy Office of Science. The SDSS-III web site is http://www.sdss3.org/.
SDSS-III is managed by the Astrophysical Research Consortium for the Participating Institutions of the SDSS-III Collaboration including the University of Arizona, the Brazilian Participation Group, Brookhaven National Laboratory, Carnegie Mellon University, University of Florida, the French Participation Group, the German Participation Group, Harvard University, the Instituto de Astrofisica de Canarias, the Michigan State/Notre Dame/JINA Participation Group, Johns Hopkins University, Lawrence Berkeley National Laboratory, Max Planck Institute for Astrophysics, Max Planck Institute for Extraterrestrial Physics, New Mexico State University, New York University, Ohio State University, Pennsylvania State University, University of Portsmouth, Princeton University, the Spanish Participation Group, University of Tokyo, University of Utah, Vanderbilt University, University of Virginia, University of Washington, and Yale University.

Financial support has been provided by project AYA2010-21887-C04-03 (former \emph{Ministerio de Ciencia e Innovaci\'on}, Spain), AYA2013-47742-C4-3-P (\emph{Ministerio de Econom\'{i}a y Competitividad}), and the `Study of Emission-Line Galaxies with Integral-Field Spectroscopy' (SELGIFS) programme, funded by the EU (FP7-PEOPLE-2013-IRSES-612701).
YA is also supported by the \emph{Ram\'{o}n y Cajal} programme (RyC-2011-09461), currently managed by the \emph{Ministerio de Econom\'{i}a y Competitividad} (still cutting back on the Spanish scientific infrastructure).

Last but not least, we are indebted to Rolf Kudritzki and I-Ting Ho for pointing out a significant typo in equation~\eqref{eq_Z_s} in an earlier version of the manuscript, and we thank Patricia S\'anchez-Bl\'azquez for useful comments, especially those regarding the stellar metallicity.

%%%%%%%%%%%%%%%%%%%%%%%%%%%%%%%%%%%%%%%%%%%%%%%%%%%%%%%%%%%%%%%%%%%%%%%%%%%%%%%
 \bibliographystyle{mn2e}
 \bibliography{references}
%%%%%%%%%%%%%%%%%%%%%%%%%%%%%%%%%%%%%%%%%%%%%%%%%%%%%%%%%%%%%%%%%%%%%%%%%%%%%%%

\end{document}